\documentclass[iop,apj,appendixfloats]{emulateapj}
\usepackage{apjfonts}
\usepackage{graphicx}
\usepackage{amsmath}
\usepackage{amssymb}

\gdef\kms{km\,s$^{-1}$}
\gdef\msun{$M_{\odot}$}

\lefthead{van Dokkum et al.}
\righthead{}
\slugcomment{Accepted for publication in ApJ Letters}
\begin{document}

\title{A high stellar velocity dispersion and $\sim 100$
globular clusters for the Ultra Diffuse Galaxy Dragonfly~44}


\author{Pieter van Dokkum\altaffilmark{1}, Roberto Abraham\altaffilmark{2},
Jean Brodie\altaffilmark{3}, Charlie
Conroy\altaffilmark{4}, Shany
Danieli\altaffilmark{1}, Allison Merritt\altaffilmark{1},
Lamiya Mowla\altaffilmark{1}, Aaron Romanowsky\altaffilmark{3,5}, Jielai
Zhang\altaffilmark{2}
\vspace{8pt}}

\altaffiltext{1}
{Astronomy Department, Yale University, New Haven, CT 06511, USA}
\altaffiltext{2}
{Department of Astronomy \& Astrophysics, University of Toronto,
   50 St.\ George Street, Toronto, ON M5S 3H4, Canada}
\altaffiltext{3}
{University of California Observatories, 1156 High Street, Santa
Cruz, CA 95064, USA}
\altaffiltext{4}
{Harvard-Smithsonian Center for Astrophysics, 60 Garden Street,
Cambridge, MA, USA}
\altaffiltext{5}
{Department of Physics and Astronomy, San Jos\'e State University,
San Jose, CA 95192, USA}

\begin{abstract}
Recently a population of large, very low surface brightness, spheroidal
galaxies was identified in the Coma cluster. The apparent survival of
these Ultra Diffuse Galaxies (UDGs) in a rich
cluster suggests that they have very high masses.
Here we present the stellar kinematics of Dragonfly~44, one of the
largest Coma UDGs,
using a 33.5\,hr integration with DEIMOS on the Keck\,II telescope. 
We find a
velocity dispersion of $\sigma=47^{+8}_{-6}$\,\kms,
which implies a dynamical mass
of $M_{\rm dyn}(<r_{1/2}) = 0.7^{+0.3}_{-0.2} \times
10^{10}$\,M$_{\odot}$ within
its deprojected half-light radius of $r_{1/2}=4.6\pm 0.2$\,kpc.
The
mass-to-light ratio is $M/L_I(<r_{1/2}) =
48^{+21}_{-14}$\,M$_{\odot}$/L$_{\odot}$, 
and the dark matter fraction is $98\,\%$
within $r_{1/2}$. 
The high  mass of Dragonfly~44 is accompanied by a large
globular cluster population.
From deep Gemini imaging taken
in $0\farcs 4$ seeing we infer that Dragonfly~44 has
$94^{+25}_{-20}$ globular clusters, 
similar to
the counts for other galaxies in this mass range.
Our results add to other recent evidence that many
UDGs are ``failed'' galaxies, with the sizes, dark matter content,
and globular cluster systems of much more luminous objects.
We estimate the total dark halo mass of Dragonfly~44 by comparing
the amount of dark matter
within $r=4.6$\,kpc to enclosed mass profiles of NFW halos. The
enclosed mass suggests a total mass of $\sim 10^{12}$\,\msun,
similar to the mass of the Milky Way.
The existence of 
nearly-dark objects with this mass is unexpected, as
galaxy formation is thought to be maximally-efficient in this
regime.

\end{abstract}

\keywords{galaxies: clusters: individual (Coma) ---
galaxies: evolution --- galaxies: structure}

\section{Introduction}
Deep imaging of the Coma cluster with the Dragonfly Telephoto
Array ({Abraham} \& {van Dokkum} 2014) uncovered a substantial
population of intrinsically-large,
very low surface brightness galaxies ({van Dokkum} {et~al.} 2015a).
These Ultra Diffuse Galaxies (UDGs) have central surface brightnesses
$\mu(g,0)>24$\,mag\,arcsec$^{-2}$ and projected half-light
radii $R_e>1.5$\,kpc. UDGs are fairly red, relatively round, and featureless;
visually, and in their central surface brightness, they
resemble dwarf spheroidal galaxies such as Sculptor and Draco,
except that their half-light radii are more than an
order of magnitude larger. Individual examples of such galaxies
had been known for many years ({Impey}, {Bothun}, \& {Malin} 1988; {Dalcanton} {et~al.} 1997), but their
ubiquity, at least in dense environments
({van Dokkum} {et~al.} 2015a; {Koda} {et~al.} 2015; {van der Burg}, {Muzzin}, \&  {Hoekstra} 2016; {Roman} \& {Trujillo} 2016), had not been recognized.

It is not clear how UDGs are related
to other classes of galaxies. One possibility is that they
are the result of processing by the cluster environment, and either
started out as small, low mass galaxies
or as very extended,
low surface brightness disks
(see, e.g., {Moore} {et~al.} 1996; {Hayashi} {et~al.} 2003; {Gnedin} 2003; {Collins} {et~al.} 2013; {Yozin} \& {Bekki} 2015).
It has been suggested that tides
were responsible for creating some of the largest and faintest galaxies
in the Local Group  ({Collins} {et~al.} 2013), and these processes are expected
to be particularly effective in clusters
({Moore} {et~al.} 1996; {Yozin} \& {Bekki} 2015).
Another idea is that UDGs represent the most rapidly-rotating tail
of the distribution of dwarf galaxies, as the size and surface brightness
of a galaxy are thought to be related to its spin
({Amorisco} \& {Loeb} 2016). The axis ratio distribution
of UDGs is inconsistent with disks under
random viewing angles ({van Dokkum} {et~al.} 2015a),
but {Amorisco} \& {Loeb} (2016) suggested that this could be
the result of processing
by the cluster environment (see also {Gnedin} 2003).

It may also be that UDGs, or a subset of them, 
are not closely related to other
low luminosity galaxies but have more
in common with galaxies that are typically much brighter. 
That is, it may be that UDGs are ``failed'' galaxies that
were prevented from building a normal stellar population,
because of extreme feedback from supernovae
and young stars ({Agertz} \& {Kravtsov} 2015; {Calura} {et~al.} 2015),
gas stripping ({Fujita} 2004; {Yozin} \& {Bekki} 2015), AGN
feedback ({Reines} {et~al.} 2013), or other effects.
Several recent studies have provided
evidence for this interpretation, as UDGs appear to
have globular cluster populations that are unusually rich
for such faint galaxies
({Beasley} {et~al.} 2016; {Peng} \& {Lim} 2016; {Beasley} \& {Trujillo} 2016). In particular, the galaxy
Dragonfly~17 in the Coma cluster has $\sim 30$ globular clusters
despite its absolute magnitude of only $M_V=-15.1$,
and could be interpreted as a ``failed'' LMC or M33
({Peng} \& {Lim} 2016; {Beasley} \& {Trujillo} 2016). These extensive globular cluster
populations suggest massive dark matter halos
({Harris}, {Harris}, \& {Alessi} 2013; {Beasley} {et~al.} 2016; {Peng} \& {Lim} 2016), and are all the more remarkable when
stripping by the cluster tidal field is taken into account
(e.g., {Smith} {et~al.} 2013).

\begin{figure*}[htb]
  \begin{center}
  \includegraphics[width=1.0\linewidth]{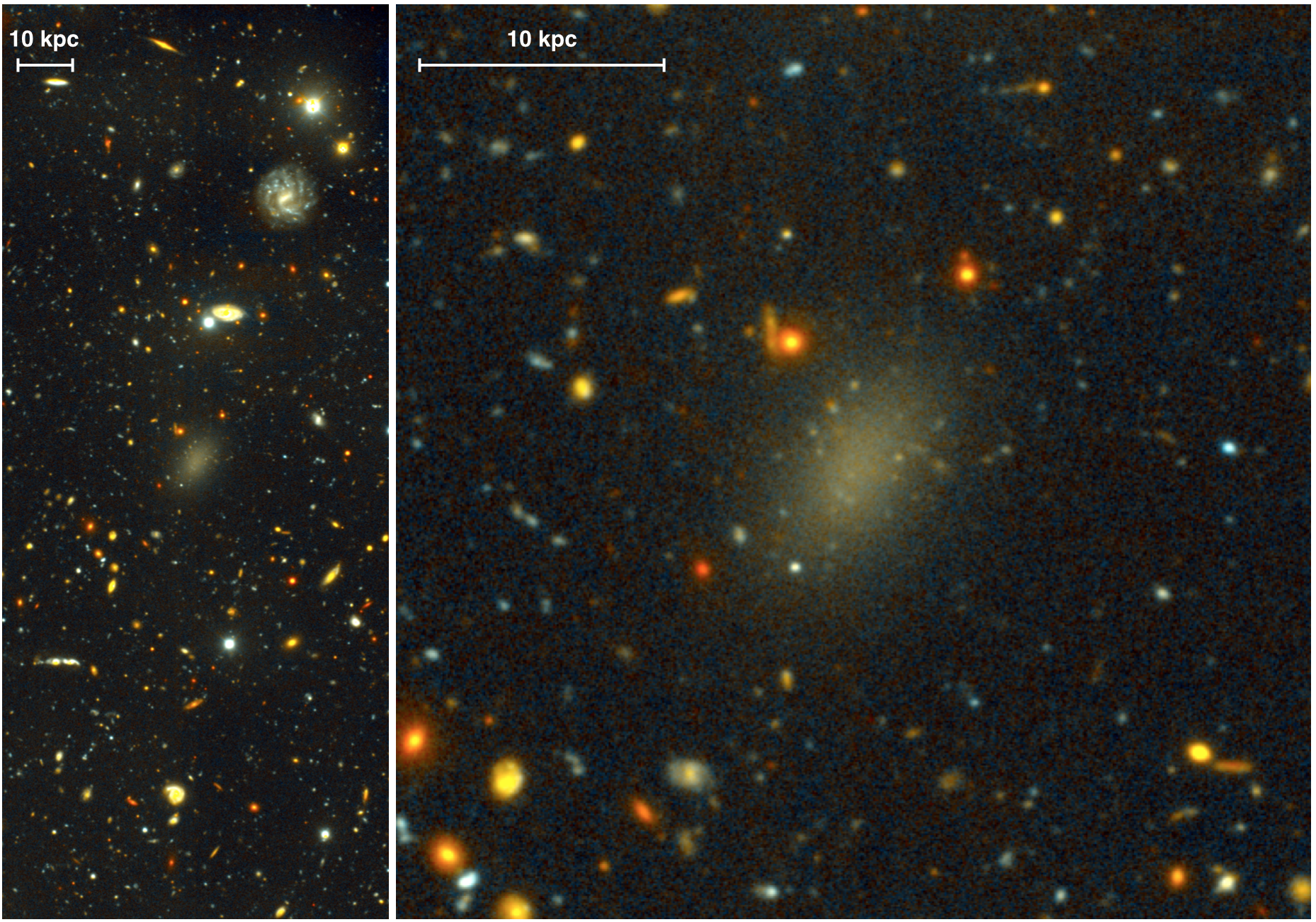}
  \end{center}
    \caption{
Deep Gemini $g$ and $i$ images were combined to create a color image
of Dragonfly~44 and its immediate surroundings. 
The galaxy has a remarkable appearance:
it is a low surface brightness, spheroidal object that is
peppered with faint, compact sources.
   }
   \label{galim.fig}
\end{figure*}

Suggestive as the globular clusters are, it is difficult to interpret
UDGs without measuring their masses.
Reliable masses are needed to
verify the assertion that the galaxies owe their structural stability
to their high dark matter fractions ({van Dokkum} {et~al.} 2015a), and can
settle the question
whether UDGs are truly distinct from other galaxies of the
same luminosity.
The first dynamical constraint on the mass of an UDG was obtained
by {Beasley} {et~al.} (2016), from the velocities of
six globular clusters attributed to
the faint ($M_g=-13.3$) galaxy VCC\,1287 in the Virgo cluster.
The velocity dispersion of
$\sigma=33^{+16}_{-10}$\,\kms\ suggests a halo mass of $\sim 10^{11}$\,\msun,
although its large uncertainty
leaves room for a range of interpretations (see {Amorisco} \& {Loeb} 2016).

In this {\em Letter} we build on these previous studies
with a measurement of the stellar
dynamics of a UDG, based on extremely deep spectroscopy with the
Deep Imaging Multi-Object Spectrograph (DEIMOS)
on the Keck\,II telescope. We also provide a measurement of the globular
cluster system of the galaxy, using ground-based imaging of exceptional
quality obtained with the Gemini-North telescope. A distance of 101\,Mpc
is assumed.

\section{Stellar Velocity Dispersion}

\subsection{Target Selection and Observations}

Dragonfly~44 is the second-largest of the 47 UDGs that were found in our
survey of the Coma cluster with the Dragonfly Telephoto Array.
Morphologically it is similar to other UDGs.
It is the only Coma UDG that has been
spectroscopically-confirmed
as a cluster member ({van Dokkum} {et~al.} 2015b), and one of only four
UDGs that have a redshift from absorption lines.\footnote{The others
are two galaxies in {Dalcanton} {et~al.} (1997) and the possible field
UDG DGSAT1 ({Mart{\'{\i}}nez-Delgado} {et~al.} 2016).}

We obtained new imaging data for Dragonfly~44, using the Gemini-North
telescope. The galaxy was observed 
on May 12, 2016 with the Gemini Multi-Object Spectrometer (GMOS)
for a total of 3000\,s in the $g$-band  and 3000\,s in the $i$-band.
Conditions were excellent, and the delivered image quality is superb:
the seeing is $0.45''$ in $g$ and $0.40''$ in $i$. 
The data were reduced using standard techniques, making use of the tasks
in the IRAF Gemini package.
A color image of the galaxy
and its immediate surroundings
is shown in Fig.\ \ref{galim.fig}.
There are no detected tidal features or other irregularities;
previously reported
variations in ellipticity
({van Dokkum} {et~al.} 2015b) can be ascribed to compact sources 
(likely globular clusters,
as discussed in Sect.\ \ref{globs.sec}) that were not recognized and masked
in the earlier, relatively poor-seeing, data.


In order to measure the galaxy's kinematics
we observed it with the DEIMOS
spectrograph on Keck II, using the 1200\,lines\,mm$^{-1}$ grating.
The slit width was
$1\farcs 0$
and the spectral resolution, as measured from sky emission lines,
is $\sigma_{\rm instr}=32$\,km\,s$^{-1}$ near the redshifted H$\alpha$ line.
The observations were carried out on January 15-16, March 11-12, and
April 9-10 2016, for a total integration time of 120,600\,s
(33.5 hrs). Conditions were excellent throughout.
The central wavelength was $\sim 6300$\,\AA.
Besides our main target three other UDGs fit in the
multi-slit mask. One of these is the faint UDG Dragonfly~42; the other
two were visually selected from archival CFHT imaging
of the Coma cluster. Results for these three galaxies will be described
elsewhere.

\begin{figure}[htb]
  \begin{center}
  \includegraphics[width=0.95\linewidth]{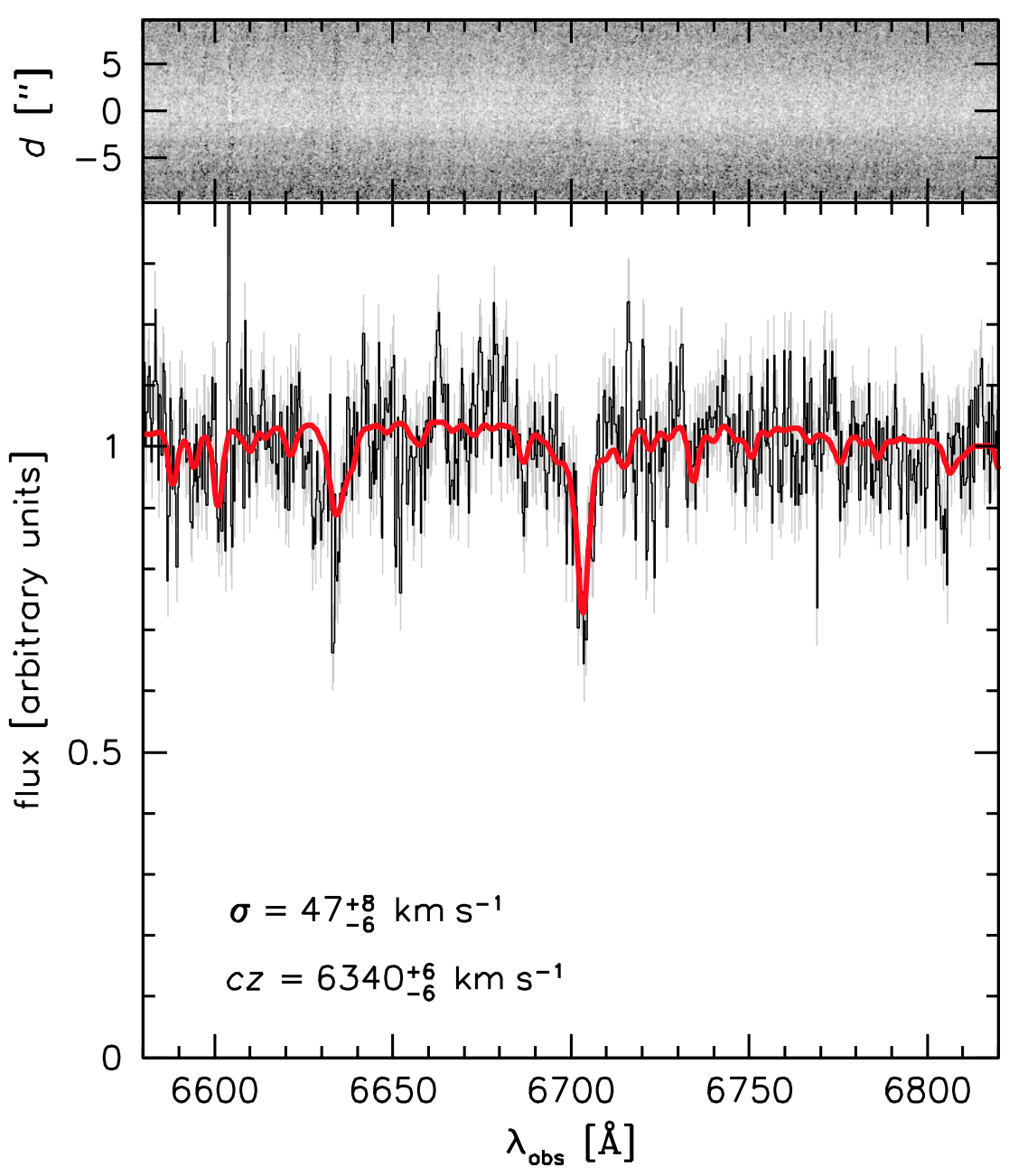}
  \end{center}
    \caption{Deep (33.5\,hr) spectrum of Dragonfly~44 obtained
with DEIMOS on the Keck II telescope. The top
panel shows the 2D spectrum. The main panel is the
collapsed 1D spectrum, with the $1\sigma$ uncertainties indicated
in grey. A flexible
model was fitted to the spectrum to determine the stellar velocity
dispersion. The best-fitting
model, with a dispersion $\sigma=47^{+8}_{-6}$\,\kms, is shown in red.
   }
   \label{spectrum.fig}
\end{figure}

We developed a custom pipeline that is optimized for faint spatially-extended
objects. Differences with the widely-used DEEP2 pipeline ({Cooper} {et~al.} 2012) include
a full modeling and subtraction of cross-talk; the use of sky lines
rather than arc lines to create the distortion model; and a careful
treatment of the background
to avoid subtracting light from the large, diffuse targets during the
reduction. 
The 2D spectrum and the collapsed 1D spectrum
are shown in Fig.\ \ref{spectrum.fig}.
The signal to noise ratio is
$14$ per 0.32\,\AA\ pixel, corresponding to ${\rm S}/{\rm N}=21$ per resolution
element. The dominant feature 
is the redshifted H$\alpha$ absorption line.

\subsection{Velocity Dispersion Measurement}

The velocity dispersion was
determined in the wavelength region 6580\,\AA\,$<\lambda<$\,6820\,\AA.
The spectrum was fitted with high resolution stellar population
synthesis models  ({Conroy}, {Gunn}, \& {White} 2009),
using
an implementation of the {\tt emcee} Markov chain Monte Carlo
sampler ({Foreman-Mackey} {et~al.} 2013) to provide reliable errors that take parameter
correlations into account.
The fit finds the best linear combination
of three templates, explicitly marginalizing over age and
metallicity, and uses both multiplicative and additive
polynomials to filter the continuum. After dividing them by the
formal errors the residuals
from the best fit have
an rms scatter of $\approx 1.0$, which
shows that the formal uncertainties
correctly describe the true errors in the data.

We find a stellar line-of-sight
velocity dispersion of $\sigma=47^{+8}_{-6}$\,km\,s$^{-1}$.
The uncertainty in $\sigma^2$, which enters the dynamical mass,
is 0.13\,dex, considerably smaller than the uncertainty of 0.30\,dex
achieved for VCC\,1287 by Beasley et al.\ (2016).
There is no evidence for rotation; any systematic trend over $\pm 5''$
is $\Delta v < 10$\,km\,s$^{-1}$, which implies that Dragonfly~44
is dispersion-dominated with $v/\sigma \lesssim 0.2$. There is also
no evidence for radial variation in the velocity dispersion.
To test the robustness of the best-fit dispersion we varied the
templates and continuum filtering; masked the
H$\alpha$ line in the fit; split the data in four independent sets
(the January run, the March run, and the two nights of the April
run) and fitted those independently; and split the data in five spatial
bins and fitted those independently. In all cases the best-fit dispersion
(or the error-weighted combination of the independent fits) is well within
$1\sigma$ of our default value of $47^{+8}_{-6}$\,\kms.

\section{Mass and Mass-to-light Ratio Inside $r_{1/2}$}

We combine the
velocity dispersion with the projected half-light
radius $R_e$ to determine the dynamical
mass and mass-to-light ($M/L$) ratio of
Dragonfly~44.
We re-measured the half-light radius
of Dragonfly~44 using the co-added $g+i$
Gemini image.
A 2D Sersic fit ({Peng} {et~al.} 2002)
gives $R_e = 8.7'' \pm 0.3''$ (4.3\,kpc
at the distance of the Coma cluster), a Sersic index $n=0.85$,
and an axis ratio $b/a=0.66$. These results are fully consistent with
previous measurements for this galaxy ({van Dokkum} {et~al.} 2015a, 2015b).
The circularized projected half-light radius
$R_{e,c}=R_e\times \sqrt{b/a}=3.5$\,kpc,
and the deprojected 3D circularized half-light radius
$r_{1/2} \approx 4/3  R_{e,c} = 4.6\pm 0.2$\,kpc ({Wolf} {et~al.} 2010).

\begin{figure*}[htb]
  \begin{center}
  \includegraphics[width=1.00\linewidth]{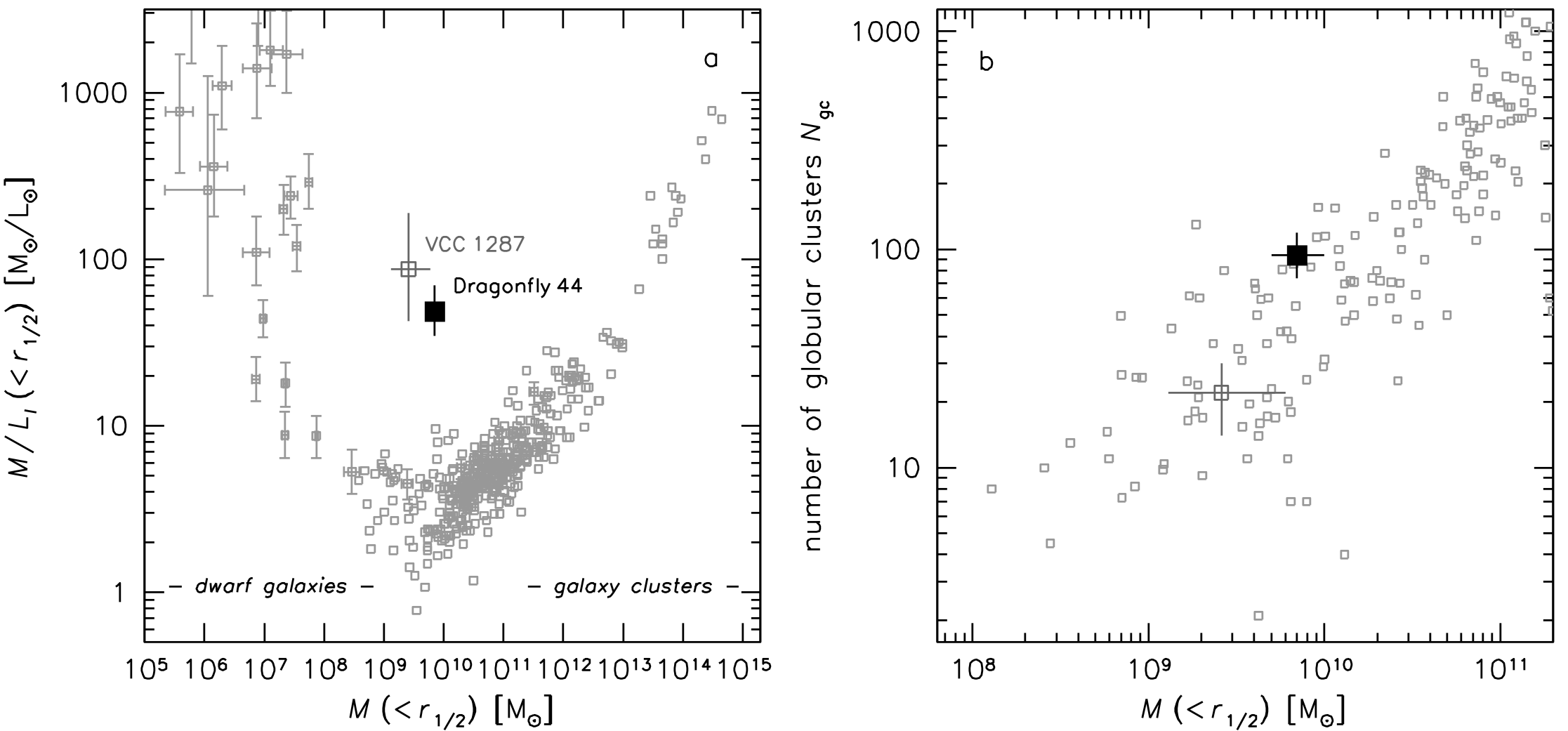}
  \end{center}
    \caption{
{\em a)} 
Relation between dynamical $M/L_I$ ratio and dynamical mass.
Open symbols are dispersion-dominated
objects from {Zaritsky}, {Gonzalez}, \&  {Zabludoff} (2006) and {Wolf} {et~al.} (2010).
The UDGs
VCC\,1287 ({Beasley} {et~al.} 2016) and Dragonfly~44 fall outside of the
band defined by the other galaxies,
having a very high $M/L$ ratio for their 
mass.  {\em b)}
Relation between the number of globular clusters $N_{\rm gc}$ and
dynamical mass. Open symbols are from the {Harris} {et~al.} (2013) compilation.
The UDGs are consistent with the relation
defined by other galaxies in this luminosity-independent plane.
   }
   \label{ml.fig}
\end{figure*}


For dynamically-hot systems 
the luminosity-weighted stellar velocity dispersion,
combined with the projected half-light radius $R_e$,
strongly constrains the
mass within the 3D half-light radius $r_{1/2}$:
\begin{equation}
M(r<r_{1/2}) \approx 9.3\times10^5 \sigma^2 R_e,
\end{equation}
with $M$ in M$_{\odot}$,
$\sigma$ in km\,s$^{-1}$ and $R_e$ in kpc ({Wolf} {et~al.} 2010).
We find $M(r<r_{1/2})=0.71^{+0.26}_{-0.17}\times 10^{10}$\,M$_{\odot}$.


This mass is much higher than expected from the stellar population alone.
Scaling the GALFIT model to the well-calibrated CFHT images
of the galaxy (see {van Dokkum} {et~al.} 2015a) and transforming from
$g$ and $i$ to $V$ and $I$, we find total magnitudes of $M_V=-16.08$
and $M_I=-17.11$ for Dragonfly~44. 
The mass-to-light ratio within $r_{1/2}$ is $M/L_I(r<r_{1/2}) =
48^{+21}_{-14}\,{\rm M}_{\odot}/{\rm L}_{\odot}$. As shown
in Fig.\ \ref{ml.fig}a, such high
$M/L$ ratios within the half-light radius are typical for very low mass
dwarf galaxies and for galaxy clusters, but not for dispersion-dominated
galaxies with
the mass of Dragonfly~44.

We calculate the dark matter fraction inside $r_{1/2}$ explicitly
by assuming that the gas fraction is negligible.
The stellar mass of Dragonfly~44, as determined from
its $i$-band luminosity and $g-i$ color ({Taylor} {et~al.} 2011), is
$M_* \approx 3\times 10^8$\,M$_{\odot}$. Therefore, the dark
matter fraction inside $r_{1/2}$ is $f_{\rm dm}=(M(r<r_{1/2})-0.5 M_*)
/M(r<r_{1/2}) \approx 98$\,\%. This amount of dark matter is sufficient
to prevent disruption of the galaxy by the Coma tidal field, at least
at distances $\gtrsim 100$\,kpc from the center of the
cluster ({Gnedin} 2003; {van Dokkum} {et~al.} 2015a; {van der Burg} {et~al.} 2016).



\begin{figure*}[htb]
  \begin{center}
  \includegraphics[width=0.75\linewidth]{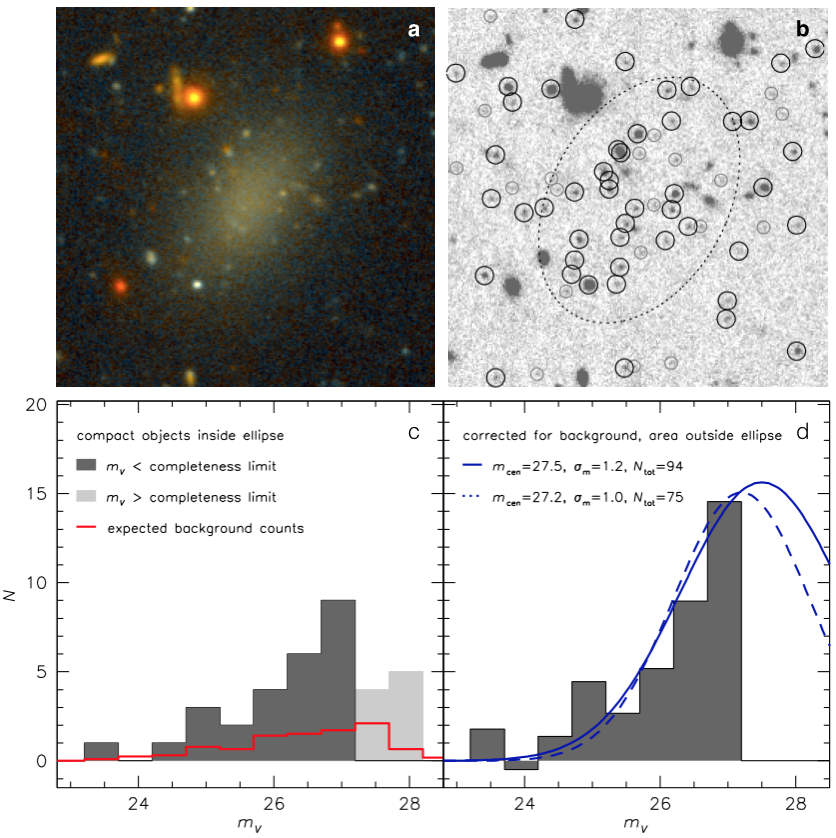}
  \end{center}
    \caption{
{\em a)}
Enlargement of the color image shown in Fig.\ 1. {\em b)}
Summed $g$ and $i$ image, after subtracting a 2D
model for the galaxy. Black, large circles indicate compact objects brighter
than the completeness limit. Grey circles
are fainter objects. The broken ellipse indicates the
assumed half-number semi-major axis of the globular clusters: $R_{\rm gc}
=1.5 R_e=6.5$\,kpc. {\em c)}
agnitude distribution of compact sources
with $R<R_{\rm gc}$. The red curve indicates the expected contribution from
background objects.
{\em d)}
agnitude distribution brighter than the completeness limit,
after subtracting the expected background and multiplying by two to
include objects with $R>R_{\rm gc}$. The blue curves are
fits to the distribution for different assumptions for the turnover
magnitude. For the expected turnover
$m_{\rm cen}=27.5$, the total number of globular clusters
is $N_{\rm gc}=94^{+25}_{-20}$.
}
   \label{globs.fig}
\end{figure*}

\section{Globular Clusters}
\label{globs.sec}


The high dynamical mass of Dragonfly~44 is accompanied by a
remarkable population of compact sources, which we identify as
globular clusters (see Fig.\ 1 and Fig.\ \ref{globs.fig}a).
Figure \ref{globs.fig}b shows all compact objects with $m_V\lesssim 28$
in the combined $g+i$ image. They were identified with
SExtractor ({Bertin} \& {Arnouts} 1996) after subtracting a 2D ellipse fit to the
galaxy. Large circles indicate objects brighter than the 80\,\%
completeness limit of $m_V=27.2$.

The spatial distribution of the globular clusters is broadly
similar to that of the galaxy light (see Fig.\ \ref{globs.fig}a), and
we measure the number of compact objects within an ellipse that
has the same orientation and axis ratio as the galaxy.
It is well
established that globular clusters have a more extended distribution
than a galaxy's stellar light ({Kartha} {et~al.} 2014), and we assume
that the half-``number'' radius
is $R_{\rm gc}=1.5\times R_e=6.5$\,kpc. This is a somewhat
conservative estimate: well-studied luminous galaxies have
$R_{\rm gc} \sim 1.8\times R_e$  ({Kartha} {et~al.} 2014), 
and  Dragonfly~17 has 
$R_{\rm gc} \sim 1.7\times R_e$ ({Peng} \& {Lim} 2016).
We find 35 compact objects within $R_{\rm gc}$, 26 of which have
$m_V<27.2$ (Fig.\ \ref{globs.fig}c). The red histogram in Fig.\
\ref{globs.fig}c is the expected magnitude distribution of unrelated
compact objects, based on empty regions in the Gemini image. 
The background-corrected number of compact objects with $R<R_{\rm gc}$
and $m_V<27.2$ is 19.3, or 38.5 when including objects outside
$R=R_{\rm gc}$. 

The luminosity function of globular clusters is well-approximated
by a Gaussian ({Harris} {et~al.} 2013), with a
turnover magnitude of
$m_{V,\rm cen} \approx 27.5$ at the distance of Coma
({Miller} \& {Lotz} 2007; {Lee} \& {Jang} 2016; {Peng} \& {Lim} 2016).
With 38.5 globular clusters having $m_V<27.2$, we derive
a total population of $N_{\rm gc}=94^{+25}_{-20}$ (solid blue curve 
in Fig.\ \ref{globs.fig}d). The errors do not include systematic
uncertainties.
The Gemini images are sufficiently deep to provide a lower limit
of $m_{\rm cen}\approx 27.2$, and for this turnover magnitude we
derive a total population of $N_{\rm gc}\approx 75$
(dashed blue curve in Fig.\ \ref{globs.fig}d). This number is reduced further
to $N_{\rm gc}\approx 63$ if we also assume that $R_{\rm gc}=R_e$ rather
than $1.5\times R_e$.

The preferred value of $N_{\rm gc}=94$ is an
order of magnitude larger than expected
for galaxies with the luminosity of Dragonfly~44:
the expected number of globular clusters for a galaxy
with $M_V=-16.1$ is $N_{\rm gc}=8^{+14}_{-5}$, where the error bars indicate
68\,\% of the 
distribution in the {Harris} {et~al.} (2013) compilation.
The specific frequency is $S_N = N_{\rm gc}10^{0.4(M_V+15)} = 35^{+9}_{-7}$,
similar to that of VCC\,1287 and Dragonfly~17
({Beasley} {et~al.} 2016; {Peng} \& {Lim} 2016; {Beasley} \& {Trujillo} 2016).
However, as shown in Fig.\ \ref{ml.fig}b, the number of globular clusters
{\em is} similar to that of other galaxies with the same {\em mass}.
The expected number of clusters for a galaxy with
$M_{\rm dyn}(<r_{1/2})=0.7\times
10^{10}$\,\msun\ is $36^{+60}_{-23}$, which is not significantly different
from the observed number. The difference would be even smaller if we had
corrected the Harris et al.\ (2013) data points for the (large)
contribution of baryons to the mass within $r_{1/2}$.

\section{Discussion}



We have shown that 
the UDG Dragonfly~44 not only has a large size for its
luminosity, it also has an anomalously large dynamical mass
and globular cluster population.
These results effectively rule out
the hypothesis that all UDGs are
rapidly-rotating or puffed-up versions of other low luminosity galaxies
(e.g., {Amorisco} \& {Loeb} 2016).
Instead, the few UDGs that have been studied in detail
({Beasley} {et~al.} 2016; {Peng} \& {Lim} 2016, and this study)
appear to be ``failed'' equivalents of more
massive galaxies: it is their low luminosity, and the lack of a classical
disk and bulge, that is anomalous.


As noted in \S\,1 it is not yet understood what physical
processes are responsible for halting or preventing star formation in UDGs. 
As these processes, and galaxy formation in general, are
thought to be a strong function
of halo mass (e.g., {Croton} {et~al.} 2006; {Dekel} {et~al.} 2009; {Behroozi}, {Wechsler}, \&  {Conroy} 2013; {Moster}, {Naab}, \& {White} 2013),
it is important to constrain
the total amount of dark matter of Dragonfly~44.
Following {Beasley} {et~al.} (2016), we estimate the halo mass by
comparing the enclosed mass within $r_{1/2}$ to cumulative
mass profiles of
theoretical models. {Beasley} {et~al.} (2016) assumed profiles from the EAGLE
simulation ({Crain} {et~al.} 2015),
which include baryons. As UDGs are completely dominated
by dark matter we use {Navarro}, {Frenk}, \& {White} (1997) profiles instead, with
a mass-dependent concentration $c$ as parameterized by {Macci{\`o}}, {Dutton}, \& {van den  Bosch} (2008). 

\begin{figure}[htb]
  \begin{center}
  \includegraphics[width=0.99\linewidth]{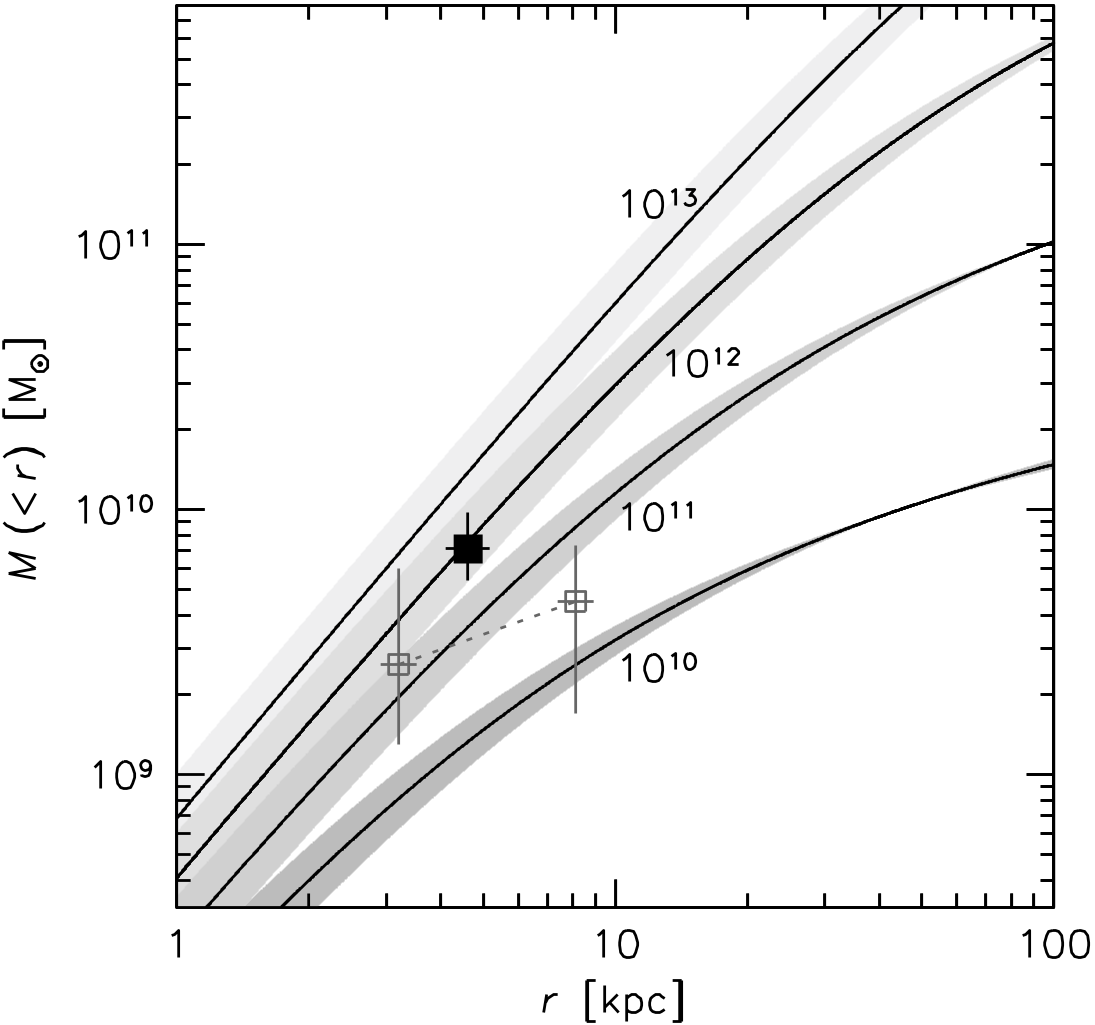}
  \end{center}
    \caption{
NFW halos ({Navarro} {et~al.} 1997)
with different masses within $r_{200}$.
The black filled square is the 
enclosed mass of Dragonfly~44 within its
deprojected half-light radius $r_{1/2}$.
The light open squares are 
for VCC\,1287; the two points are for
$r_{1/2}$ and for the
radius that includes all six
globular clusters with measured velocities (see {Beasley} {et~al.} 2016).
   }
   \label{halo.fig}
\end{figure}

The results are shown in Fig.\ \ref{halo.fig}, for halos with 
$M_{\rm halo}(<r_{200})=10^{10}-10^{13}\,{\rm M}_{\odot}$.
The grey bands
indicate the variation
in the enclosed mass profiles for a halo-to-halo scatter in concentration
of $\Delta(\log c)=0.14$, and illustrate the degeneracy between
concentration and derived halo
mass (see, e.g., {Taylor} {et~al.} 2015).
The  observed enclosed mass of Dragonfly~44 suggests
a halo mass of
$M_{\rm halo}(<r_{200})\approx 8\times 10^{11}\,{\rm M}_{\odot}$,
if the halo has an average concentration and no
truncation
(see, e.g., {Gnedin} 2003, for a discussion of these assumptions).

Therefore, whereas  VCC\,1287 (and also Dragonfly~17)
can be considered ``failed'' LMCs or M33s,
the more massive Dragonfly~44  can be viewed as
a failed Milky Way. This distinction is potentially important:
it is the accepted view that the ratio of
stellar mass to halo mass reaches a peak of $\sim 0.03$ for
$M_{\rm halo}\sim 10^{12}\,{\rm M}_{\odot}$, which suggests
that galaxy formation is maximally efficient in halos of this
mass ({Behroozi} {et~al.} 2013; {Moster} {et~al.} 2013). Dragonfly~44 has a stellar mass that is a
factor of $\sim 100$ lower than expected in this framework, and
in a standard halo abundance matching exercise it would be
assigned the wrong halo mass.\footnote{We note that this is based
on the stellar mass -- halo mass relation
for field galaxies at $z=0$; as discussed in, e.g.,
Grossauer et al.\ (2015), the discrepancy may be smaller for
cluster galaxies and if the galaxies formed at high redshift.}
More importantly, whatever physical
processes are responsible for forming galaxies such as Dragonfly~44, they
can apparently operate in a regime where galaxy formation was
thought to be both maximally-efficient and relatively well understood.


We emphasize, however, that the halo abundance matching technique
relies on total halo masses, and in our study the total halo mass is
an extrapolation
of the measured mass by a factor of $\sim 100$.
A more robust and less model-dependent conclusion
is that the dark matter mass within $r=4.6$\,kpc is similar
to the dark matter mass of the Milky
Way 
within the same radius ({Xue} {et~al.} 2008).
Better constraints on the halo masses of UDGs may come from
lensing studies of large samples. Intriguingly, a 
weak lensing map of the Coma cluster by {Okabe} {et~al.} (2014) shows
a $2\sigma$ peak 
at the location of Dragonfly~44. Peaks of similar significance
have inferred masses of a few $\times 10^{12}\,{\rm M}_{\odot}$, and
unlike most other features in the map it is not associated with known bright
galaxies or background structures.

Our study demonstrates
that it is possible to measure the
stellar kinematics of UDGs using existing instrumentation
on large telescopes. With sufficiently large samples
it will be possible to determine
what fraction of UDGs are ``failed'' galaxies (as opposed to, say,
tidally-stretched low mass galaxies), and what the variation
is in their masses and $M/L$ ratios.
A preliminary analysis of the other, smaller, UDGs in our DEIMOS mask suggests
that they have lower velocity dispersions than Dragonfly~44; a study
of the ensemble of UDGs is in preparation.





\begin{acknowledgements}
We thank the anonymous referee for insightful comments which improved
the manuscript.
\end{acknowledgements}


\end{document}